\documentclass[aps,twocolumn,floats,superscriptaddress,prd,nofootinbib]{revtex4}
\usepackage{graphicx, epsfig, bm, amsmath}

\usepackage{color}
\usepackage{hyperref}
\usepackage{ifthen}
\usepackage{xstring}

\begin{document}


\newcommand{\zmin}{z_{\rm min}}
\newcommand{\zmax}{z_{\rm max}}
\newcommand{\dz}{\Delta z}
\newcommand{\dzsub}{\Delta z_{\rm sub}}
\newcommand{\zminsn}{z_{\rm min}^{\rm SN}}
\newcommand{\nzpc}{N_{z,{\rm PC}}}
\newcommand{\nmax}{N_{\rm max}}
\newcommand{\amax}{a_{\rm max}}
\newcommand{\atr}{a_{\rm tr}}
\newcommand{\aeq}{a_{\rm eq}}

\newcommand{\wmin}{w_{\rm min}}
\newcommand{\wmax}{w_{\rm max}}
\newcommand{\wfid}{w_{\rm fid}}

\newcommand{\lcdm}{$\Lambda$CDM}

\newcommand{\thetal}{\bm{\theta}_{\Lambda}}
\newcommand{\thetaq}{\bm{\theta}_{\rm Q}}
\newcommand{\thetade}{\bm{\theta}_{\rm DE}}
\newcommand{\thetas}{\bm{\theta}_{\rm S}}
\newcommand{\thetadel}{\bm{\theta}_{{\rm DE},\Lambda}}
\newcommand{\thetadeq}{\bm{\theta}_{\rm DE,Q}}
\newcommand{\thetades}{\bm{\theta}_{\rm DE,S}}
\newcommand{\thetaother}{\bm{\theta}_{\rm nuis}}

\newcommand{\om}{\Omega_{\rm m}}
\newcommand{\ode}{\Omega_{\rm DE}}
\newcommand{\ok}{\Omega_{\rm K}}
\newcommand{\omhh}{\Omega_{\rm m} h^2}
\newcommand{\obhh}{\Omega_{\rm b} h^2}
\newcommand{\winf}{w_{\infty}}
\newcommand{\scrm}{\mathcal{M}}
\newcommand{\osf}{\Omega_{\rm sf}}
\newcommand{\omf}{\Omega_{\rm m}^{\rm fid}}
\newcommand{\scrmf}{\mathcal{M}^{\rm fid}}
\newcommand{\rhode}{\rho_{\rm DE}}
\newcommand{\rhoc}{\rho_{{\rm cr},0}}
\newcommand{\dlnl}{-2\Delta\ln\mathcal{L}}
\newcommand{\dlum}{d_{\rm L}}
\newcommand{\dlss}{D_*}

\newcommand{\zh}{z_h}
\newcommand{\zbao}{z_{\rm BAO}}

\newcommand{\gpr}{G^{\prime}}
\definecolor{darkgreen}{cmyk}{0.85,0.2,1.00,0.2}

\newcommand{\mmcomment}[1]{\textcolor{red}{[{\bf MM}: #1]}}
\newcommand{\dhcomment}[1]{\textcolor{darkgreen}{[{\bf DH}: #1]}}
\newcommand{\wh}[1]{\textcolor{blue}{[{\bf WH}: #1]}}


\pagestyle{plain}

\title{Testable dark energy predictions from current data}

\author{Michael J.\ Mortonson}
\affiliation{Center for Cosmology and AstroParticle Physics, 
        The Ohio State University, Columbus, OH 43210}
\affiliation{Department of Physics,
        University of Chicago, Chicago, IL 60637}
\affiliation{Kavli Institute for Cosmological Physics 
       and Enrico Fermi Institute,
        University of Chicago, Chicago, IL 60637}

\author{Wayne Hu}
\affiliation{Kavli Institute for Cosmological Physics 
       and Enrico Fermi Institute,
        University of Chicago, Chicago, IL 60637}
\affiliation{Department of Astronomy \& Astrophysics,
        University of Chicago, Chicago, IL 60637}

\author{Dragan Huterer}
\affiliation{Department of Physics, University of Michigan, 
450 Church St, Ann Arbor, MI 48109-1040}

\begin{abstract}
  Given a class of dark energy models, constraints from one set of cosmic
  acceleration observables make predictions for other observables.  
  Here we present
  the allowed ranges for the expansion rate $H(z)$, distances $D(z)$, and the
  linear growth function $G(z)$ (as well as other, derived growth observables) 
  from the current combination of cosmological
  measurements of supernovae, the cosmic microwave background, baryon acoustic
  oscillations, and the Hubble constant.  With a cosmological
  constant as the dark energy and assuming near-minimal neutrino masses, 
  the growth function is already predicted to
  better than $2\%$ precision at \emph{any} redshift, with or without 
  spatial curvature.  Direct measurements of growth that 
  match this precision offer the opportunity to
  stringently test and potentially rule out a cosmological constant.  While
  predictions in the broader class of quintessence models are weaker, it is
  remarkable that they are typically within a factor of $2-3$ of forecasts for
  future space-based supernovae and Planck CMB measurements.  In particular,
  measurements of growth at any redshift, or the Hubble constant $H_0$, that exceed
  $\Lambda$CDM predictions by substantially more than $2\%$ would rule out 
  not only a cosmological constant but also the whole quintessence
  class, with or without curvature and early dark energy. 
  Barring additional systematic errors hiding in the data, 
  such a discovery would require more exotic explanations of 
  cosmic acceleration such as phantom dark energy, dark energy clustering, 
  or modifications of gravity.
\end{abstract}

\maketitle

\section{Introduction}
\label{sec:intro}

Within a fixed class of dark energy models, such as the cosmological constant
or scalar field quintessence, various cosmological observables are all
interrelated by the properties of the class itself.
The narrower the class,
the higher the expected correlation between measurements of different
observables. Therefore, given a class of dark energy models, constraints from
one set of cosmic acceleration observables make predictions for other observables.
For example, it is well known that since the first release of WMAP data
\cite{Spergel_2003}, the Hubble constant in a spatially flat universe with a
cosmological constant and cold dark matter ($\Lambda$CDM) has been predicted
to a precision better than it has yet been measured.  Predictions like this one
therefore offer the opportunity for more precise measurements to falsify
the dark energy model (in this case, flat $\Lambda$CDM)~\cite{Hu_standards}.

In a previous paper (hereafter MHH) \cite{PaperI}, we showed how this idea
can be generalized to additional acceleration observables and wider classes of
dark energy models.  Other observables include the expansion rate $H(z)$, the 
comoving angular diameter distance $D(z)$, and the linear growth function
$G(z)$.  The model classes we considered include a cosmological constant, 
with and without
spatial curvature, and scalar field quintessence models, with and without 
early dark energy and spatial curvature components. 
Using forecasts for a Stage IV \cite{DETF} SN sample and Planck CMB data, we found 
that future data sets will provide numerous strong predictions that we 
may use to attempt to falsify various acceleration paradigms.

\vspace{1cm}
In this paper, we evaluate the predictive power of {\it current} measurements
to constrain the expansion rate, distance, and growth as a function of
redshift.  Specifically, we consider current measurements of supernovae (SN),
the cosmic microwave background (CMB), baryon acoustic oscillations (BAO), and
the Hubble constant ($H_0$).  These predictions 
target the redshift ranges and required precision for future measurements
seeking to rule out whole classes of models for cosmic acceleration.

Our approach complements studies that seek to constrain an ever expanding set
of parameters of the dark energy. The most ambitious analyses currently
utilize $\sim 5$ parameters to describe the dark energy equation of state
$w(z)$ \cite{Huterer_Cooray,Wang_Tegmark_2005,Riess_2006,Zunckel_Trotta,
  Sullivan_Cooray_Holz,Zhao_Huterer_Zhang,Zhao_Zhang:2009,Serra:2009}.  We
take these studies in a new direction: rather than constraining parameters
associated with the equation of state, we propagate constraints from the data
into allowed ranges for $H(z)$, $D(z)$, $G(z)$, and auxiliary observables that
can be constructed from them through a principal component representation of
$w(z)$ that is complete in these observables for $z<1.7$.  This work goes
beyond previous studies that are similar in spirit
(e.g.~\cite{Kujat,Sahlen05,Huterer_Peiris,Chongchitnan_Efstathiou,ZhaKnoTys08})
by directly applying constraints from current data sets
  to complete representations of several dark energy model classes and making
  concrete predictions for a number of observable quantities.

This paper is organized as follows.
We begin in Sec.~\ref{sec:methods} with a discussion of the methodology of
predicting observables within classes of dark energy models, including 
descriptions of each of the acceleration observables, cosmological data sets, 
and model classes.  We present our
predictions from current data in Sec.~\ref{sec:predict} 
and discuss the results in Sec.~\ref{sec:discussion}.

\section{Methodology}
\label{sec:methods}

\subsection{Acceleration Observables}
\label{sec:obs}

There are two general types of acceleration observables: those related
to the expansion history and geometry of the universe, and those related to the growth of structure.
In terms of a general evolution for the dark energy 
equation of state $w(z)$,  
the expansion history observables are the Hubble expansion rate
\begin{eqnarray}
H(z) &=& H_0 \left[ \om (1+z)^3 + \ode f(z) + \ok (1+z)^2 \right]^{1/2},
\nonumber \\
&& f(z) = \exp\left[3 \int_0^z dz' \frac{1+w(z')}{1+z'}\right], \label{eq:hz} 
\end{eqnarray}
where $\om$ and $\ode$ are the present matter and dark energy densities, 
respectively, as fractions of the critical density for flatness, 
spatial curvature is parametrized by $\ok\equiv 1-\om-\ode$, 
and the small contribution of radiation at $z\sim 1$ is neglected; 
and the comoving angular diameter distance
\begin{equation}
D(z) = \frac{1}{(|\ok|H_0^2)^{1/2}} S_{\rm K}\left[(|\ok|H_0^2)^{1/2} 
\int_0^z \frac{dz'}{H(z')} \right],
\label{eq:dist}
\end{equation}
where the function $S_{\rm K}(x)$ is equal to $x$ in a flat universe ($\ok=0$), 
$\sinh x$ in an open universe ($\ok>0$), and $\sin x$ in a closed universe ($\ok<0$).
The growth of
linear density perturbations $\delta \propto G a$
is given by
\begin{equation}
G'' + \left(4+\frac{H'}{H}\right)G' + \left[
3+\frac{H'}{H}-\frac{3}{2}\om(z)\right]G = 0,
\label{eq:growth}
\end{equation}
where primes denote derivatives with respect to $\ln a$ and $\om(z) = \om
H_0^2(1+z)^3/H^2(z)$.  We assume scales for which the dark energy density is
spatially smooth compared with the matter and normalize $G(z)=1$ at $z=10^3$.

There are several auxiliary quantities related to the growth function that are also
interesting to examine.   Since growth measurements like the evolution of the
cluster abundance often compare the change in growth relative to the present, we
also consider a different normalization for the growth function,
\begin{equation}
G_0(z)\equiv \frac{G(z)}{G(0)}.
\end{equation}
Velocity field measurements, on the other hand, are sensitive to the growth {\it rate}
\begin{equation}
f(z) \equiv  1 + \frac{G'}{G}.
\end{equation}
Specifically, the amplitude of the velocity power spectrum can be measured
from redshift space distortions and constrains $f(z)G(z)$ independently of galaxy bias (e.g.\ see~\cite{Percival_White}).
Finally, given that the growth rate is approximately related to expansion
history observables by $f(z) = [\om(z)]^{\gamma}$ where the growth index is
$\gamma \approx 0.55$ for flat \lcdm\ \cite{WangSteinhardt,Linder_gamma} we
also consider predictions for
\begin{equation}
\gamma(z)\equiv {\ln[f(z)] \over \ln[\om(z)]} \,.
\label{eq:gamma_z}
\end{equation}
Note however that $\gamma(z)$ is not a direct observable but rather 
must be inferred from a combination of measurements in a 
specific dark energy context.

We ignore the influence of massive neutrinos throughout this study.  
The effect of massive neutrinos on the growth of structure is
significantly scale-dependent, 
but on present linear scales well below the horizon, 
$k\sim 0.01-0.1~h$~Mpc$^{-1}$, the growth suppression from a 
normal neutrino mass hierarchy with $\sum m_{\nu}\sim 0.05$~eV \cite{nu_oscill}
is $\lesssim 1\%$ in $G(z)$ and $f(z)G(z)$ and smaller for other 
observables. The maximum decrement in growth from nearly-degenerate 
neutrinos with $\sum m_{\nu}\sim 0.5$~eV (e.g.~\cite{Reid_SDSSDR7}) 
is $\sim 1-10\%$ on these scales. In the predictions we present here, 
these effects would appear as an additional ``early'' dark energy 
component with $w\approx 0$. Future precise measurements of $\sum m_{\nu}$ 
from independent data could be used to correct the growth predictions here 
by scaling them by the appropriate suppression factor.

\subsection{Constraints from Current Data}
\label{sec:data}

The main observational constraints we consider when making predictions for
acceleration observables include relative distances at $z \lesssim 1.5$ from
Type Ia SNe and absolute distances at $z_*=1090$ from the CMB, $\zbao\approx
0.35$ from BAO, and $\zh\approx 0.04$ from low-redshift SNe calibrated with
maser and Cepheid distances. Since low-$z$ distances mainly probe the Hubble
constant for smoothly varying $w(z)$, we refer to the low-$z$ SN
calibration as an $H_0$ constraint.  The CMB data additionally constrain
parameters that impact dark energy models such as the matter density $\omhh$
and the fraction of dark energy density at recombination.

In the simplest classes of models, the SN and CMB data suffice to make
accurate predictions for expansion and growth observables.  In more complex
classes, BAO and $H_0$ constraints on distances are necessary.  Even
in these cases, predictive power is still retained in that measured distances
to a few specific redshifts constrain $H(z)$, $D(z)$, and $G(z)$ at all redshifts.
We now describe each of these data sets in more detail.

The Type Ia SN sample we use is the Union compilation~\cite{SCP_Union}.  These
SN observations measure relative distances, $D(z_1)/D(z_2)$, over a range of
redshifts spanning $0.015 \leq z \leq 1.551$, with most SNe at $z \lesssim
1$. We add the SN constraints using the likelihood code for the Union data
sets \cite{Union_like}, which includes estimated systematic errors for the SN
data~\cite{SCP_Union}.

For the CMB, we use the most recent, 5-year release of data from the WMAP
satellite \cite{Komatsu_2008,Nolta_2008,Dunkley_2008} employing the likelihood
code available at the LAMBDA web site \cite{WMAP_like}.  Unlike the CMB
distance priors on $D(z_*)$ and $\omhh$ used for the forecasts in MHH, the
likelihood used here contains the full information from the CMB angular power
spectra; in particular this provides sensitivity to large fractions of
early dark energy at recombination as well as information about late-time dark
energy and spatial curvature from the ISW effect without necessitating
additional priors.  We compute the CMB angular power spectra using the code
CAMB \cite{Lewis:1999bs,camb_url} modified with the parametrized
post-Friedmann (PPF) dark energy module \cite{PPF,ppf_url} to include 
 models with general dark energy equation of state evolution
where $w(z)$ may cross $w=-1$.  Note that while our {\it predictions} for
growth observables apply to scales on which dark energy is smooth relative to
matter, the CAMB+PPF code self-consistently accounts for the effects of 
scale-dependent dark energy perturbations on the CMB anisotropies.

The BAO constraint we use is based on the measurement of the correlation
function of SDSS Luminous Red Galaxies (LRGs) \cite{Eisenstein}, which
determines the distance and expansion rate at $\zbao\approx 0.35$ through the
combination $D_V(z) \equiv [z D^2(z)/H(z)]^{1/3}$.  We implement this
constraint by taking the volume average of this quantity, $\langle D_V \rangle$, over
the LRG redshifts, $0.16<z<0.47$, and comparing with the value of $A \equiv
\langle D_V \rangle \sqrt{\omhh}/\zbao$ given in Ref.~\cite{Eisenstein}, $A =
0.472 \pm 0.017$ (taking the scalar spectral tilt to be 
$n_s=0.96$).  We discuss the
expected impact of more recent BAO measurements \cite{Percival09} on our
predictions in Sec.~\ref{sec:discussion}.

Finally, we include the recent Hubble constant constraint from the SHOES team
\cite{SHOES}, based on SN distances at $0.023<z<0.1$ that are linked to a
maser-determined absolute distance using Cepheids observed in both the maser
galaxy and nearby galaxies hosting Type Ia SNe.  The SHOES measurement
determines the absolute distance to a mean SN redshift of $\zh=0.04$,
which effectively corresponds to a constraint on $H_0$ for models with
relatively smooth dark energy evolution in the recent past such that $\lim_{z\to
  0} D(z) = cz/H_0$.  Sharp transitions in the dark energy density at 
ultra-low redshifts can break
the relationship between low-redshift distances and $H_0$ as described in
Ref.~\cite{HubTrans}, but the principal component parametrization we use is
constructed to largely eliminate such possibilities (see MHH, Appendix B).
Nonetheless, given that the observations relate distance and redshift, and
distances are more robust to variations in the equation of state at low
redshift than is the instantaneous expansion rate, we implement the $H_0$
constraint as a measurement of $D(\zh) = c\zh/(74.2 \pm 3.6$ km~s$^{-1}$~Mpc$^{-1}$).

\subsection{Model Classes}
\label{sec:pcs}

Our basic model classes are (1) ``\lcdm,'' where dark energy is a
cosmological constant $\Lambda$ with equation of state $w=-1$, 
and (2) ``quintessence,'' the general
class of scalar field models with arbitrary but bounded equation of state
evolution $-1\leq w(z)\leq 1$.  For these two cases we maintain a {\it complete} description
of the observable degrees of freedom.
Finally, there is (3) ``smooth dark energy" which is the generalization of quintessence
to unbounded $w(z)$, assuming that dark energy is unclustered relative to matter.   
Unlike the forecasts in MHH, we do not 
maintain completeness for smooth dark energy but rather take a fixed functional form
$w(z) = w_0 + (1-a) w_a$.  This choice allows us to simply identify observables that could
potentially falsify quintessence in favor of smooth dark energy 
but does not allow us to make predictions that could falsify 
the broader class as a whole. It also allows us to identify how 
predictions in the quintessence class change if we require smooth, 
monotonic evolution in $w(z)$.   In each case, the model class can either be restricted to
spatially flat cosmologies or allow spatial curvature, parametrized by $\ok$

\begin{table*}
\caption{Dark energy model classes, their defining parameter sets and priors, 
and figures in which predictions appear.}
\begin{center}
\begin{tabular*}{\textwidth}{@{\extracolsep{\fill}}lccr}
\hline
\hline
Model Class & Parameters & Priors & Figures \\
\hline
flat \lcdm\ & $\thetal$ & $\ok=0$ & 1, 2, 3, 8 \\
non-flat \lcdm\ & $\thetal$ & none  & 2 \\
flat PC quintessence without early dark energy & $\thetaq$ & $\{\alpha_i\}$ priors\footnote{Conservative quintessence priors on PC amplitudes; see Sec.~\ref{sec:pcs}.}, $\ok=0$, $\winf=-1$ & 3, 4, 5, 6, 8 \\
flat PC quintessence with early dark energy & $\thetaq$ & $\{\alpha_i\}$ priors, $\ok=0$ & 5 \\
non-flat PC quintessence without early dark energy & $\thetaq$ & $\{\alpha_i\}$ priors, $\winf=-1$ & 6, 7 \\
non-flat PC quintessence with early dark energy & $\thetaq$ & $\{\alpha_i\}$ priors & 7, 8 \\
flat $w_0-w_a$ with quintessence priors & $\thetas$ & $-1\leq w_0\leq 1$, $-1\leq w_0+w_a\leq 1$, $\ok=0$ & 9 \\
non-flat $w_0-w_a$ with quintessence priors & $\thetas$ & $-1\leq w_0\leq 1$, $-1\leq w_0+w_a\leq 1$ & 10 \\
flat $w_0-w_a$ (no $w$ prior; smooth dark energy) & $\thetas$ & $\ok=0$ & 9 \\
non-flat $w_0-w_a$ (no $w$ prior; smooth dark energy) & $\thetas$ & none & 10 \\
\hline
\hline
\end{tabular*}
\end{center}
\label{tab:modelclasses}
\end{table*}

For the quintessence class, we follow 
 the procedure described in MHH and parametrize $w(z)$ at $z<\zmax = 1.7$ 
with a basis of principal components (PCs)
\cite{Hu_PC,Huterer_Starkman}.  For our purposes, the PCs simply act as an intermediate
basis to represent observables, required to be complete for
arbitrary variations in $w(z)$ at $z<z_{\rm max}$.  We construct the PCs
 using the specifications of
a Stage IV SN experiment, specifically the SuperNova/Acceleration Probe (SNAP)
\cite{SNAP}, combined with CMB information from the recently
launched Planck satellite.  

Specifically,
the principal component functions $e_{i}(z_{j})$ are eigenvectors of the
SNAP+Planck covariance matrix for the equation of state in redshift bins $z_j$, and
they form a basis in which an arbitrary function $w(z_j)$ may be expressed as
\begin{equation}
w(z_j) - \wfid(z_j) = \sum_{i=1}^{\nzpc} \alpha_i e_i(z_j),
\label{eq:pcstow}
\end{equation}
where $\alpha_i$ are the PC amplitudes, $\nzpc = 1+\zmax/\dz$ is the number of
redshift bins of width $\dz$, and $z_j = (j-1) \dz$. The maximum redshift for
variations in $w(z)$ ($\zmax=1.7$) matches the largest redshift for the SNAP
supernova data, and we use a fiducial model $\wfid(z)=-1$ since \lcdm\ is an
excellent fit to current data.

Since the highest-variance PCs correspond to modes of $w(z)$ to which 
both data and predicted observables are insensitive, we truncate the sum in Eq.~(\ref{eq:pcstow}) 
with $\nmax<\nzpc$ PCs.
As shown in MHH, for our choices of $\zmax$ and $\wfid(z)$, the 10
lowest-variance PCs ($\nmax=10$) form a basis which, 
for the classes of models we consider here,  is sufficiently complete 
for future Stage IV measurements and so
more than suffice for the current data.  We have also explicitly checked that
there is little difference in predictions between $\nmax=5$ and $\nmax=10$ 
for one of the
model classes, flat quintessence without early dark energy.

Quintessence models describe dark energy as a scalar field with kinetic and potential
contributions to energy and pressure.   Barring models where large kinetic
and (negative) potential contributions cancel, quintessence equations of state are 
restricted to 
$-1\leq w(z)\leq 1$.  Following MHH, this bound is conservatively implemented
with uncorrelated top-hat priors on the PC amplitudes $\alpha_i$.  
Any combination of PC amplitudes that is
rejected by these priors must arise from an equation of state $w(z)$ that
violates the bound on $w(z)$, but not all models that are allowed by the priors
strictly satisfy this bound; the set of models we consider is
therefore ``complete'' but not ``pure.''  This ambiguity arises since we
truncate the principal components at $\nmax=10$ and we wish to allow for the
possibility that the omitted components may conspire to satisfy the bound.
For the purposes of falsifying dark energy model classes a complete but impure
sampling of quintessence models is sufficient, although more efficient
rejection of models that violate the $-1\leq w(z)\leq 1$ bound could result in
somewhat tighter observable predictions \cite{Samsing_Linder}.  Further
details on the construction of the PCs and implementation of the priors can be
found in MHH.

The above prescription only includes dark energy variations at the relatively 
late times that are probed by SN data, $z<\zmax$.
To describe ``early dark energy'' at $z>\zmax$, we adopt a simple parametrization by 
assuming a constant equation of state, $w(z>\zmax)=\winf$, 
restricted to $-1\leq \winf\leq 1$.  The dark energy density at $z>\zmax$ can be
extrapolated from its value at $\zmax$ as
\begin{equation}
\rhode(z) = \rhode(\zmax)\left(\frac{1+z}{1+\zmax}\right)^{3(1+\winf)}.
\end{equation}
For more restricted model classes where we assume that there is 
no significant early dark energy, 
we fix $\winf=-1$ since a constant dark energy density 
rapidly becomes negligible relative to the matter density 
at increasing redshift.
Note that the possibility of early dark energy is automatically 
included in the smooth $w_0-w_a$
model class where the equation of state at high redshift is $w\approx w_0+w_a$.

In addition to the dark energy parameters described above ($\thetade$), we include
cosmological parameters that affect the CMB angular power spectra but 
not the acceleration observables ($\thetaother$): the physical baryon density $\obhh$, 
the normalization and tilt of the primordial curvature
spectrum $\Delta_{\cal R}^2 = A_s (k/k_0)^{n_s-1}$ with $k_0 = 0.05$~Mpc$^{-1}$,
and the optical depth to reionization $\tau$.
This brings our full set of parameters for \lcdm\ to $\thetal=\thetadel+\thetaother$,
and for quintessence and smooth $w_0-w_a$ dark energy
we define the analogous parameter sets with
\begin{eqnarray}
\thetadel &=&\{\omhh, \om, \ok\} \,,\nonumber\\
\thetadeq &=& \thetadel + \{\alpha_1,\ldots, \alpha_{\nmax}, \winf\} \,,\nonumber\\
\thetades &=& \thetadel + \{w_0, w_a\} \,,\nonumber\\
\thetaother &=& \{ \obhh, n_s, A_s, \tau \} \,,
\label{eq:parametersfull}
\end{eqnarray}
where we count $\Omega_{\rm m}$ and $\Omega_{\rm K}$ as dark energy parameters
since $\Omega_{\rm DE}= 1-\Omega_{\rm m} - \Omega_{\rm K}$.
Note that the Hubble constant is a derived parameter, $h=H_0/(100~{\rm km~s}^{-1}{\rm
  Mpc}^{-1}) = (\omhh/\om)^{1/2}$. 
Although the observable predictions mainly depend on constraints on the 
dark energy parameters $\thetade$, we include the additional ``nuisance'' 
parameters $\thetaother$ due to degeneracies between 
$\thetade$ and $\thetaother$ parameters in current CMB data; 
these nuisance parameters are marginalized over in our predictions 
for acceleration observables.
The parameter sets and priors on the parameters for each model 
class are summarized in Table~\ref{tab:modelclasses}.

\subsection{MCMC Predictions}
\label{sec:mcmc}

To make predictions for the acceleration observables using constraints from current data, we
use a Markov Chain Monte Carlo (MCMC) likelihood analysis.  Given a dark energy
model class parametrized by $\thetal$,  $\thetaq$, or $\thetas$, the MCMC algorithm
estimates the joint posterior distribution of cosmological parameters and
predicted observables by sampling the parameter space and evaluating the
likelihood of each proposed model compared with the data described in
Sec.~\ref{sec:data}
(e.g.\ see~\cite{Christensen:2001gj,Kosowsky:2002zt,Dunetal05}).  We use the
code CosmoMC \cite{Lewis:2002ah,cosmomc_url} for the MCMC analysis.

The posterior distribution is obtained using Bayes' Theorem,
\begin{equation}
{\cal P}(\bm{\theta}|{\bf x})=
\frac{{\cal L}({\bf x}|\bm{\theta}){\cal P}(\bm{\theta})}{\int d\bm{\theta}~
{\cal L}({\bf x}|\bm{\theta}){\cal P}(\bm{\theta})},
\label{eq:bayes}
\end{equation}
where ${\cal L}({\bf x}|\bm{\theta})$ is the likelihood of the data ${\bf x}$
given the model parameters $\bm{\theta}$ and ${\cal P}(\bm{\theta})$ is the
prior probability density. The MCMC algorithm generates random draws from the
posterior distribution that are fair samples of the likelihood surface.
We test convergence of the samples to a stationary distribution that
approximates the joint posterior density ${\cal P}(\bm{\theta}|{\bf x})$ 
by applying a conservative Gelman-Rubin criterion \cite{gelman/rubin}
of $R-1\lesssim 0.01$ across a minimum of four chains for each model class.

As described in MHH, the MCMC approach allows us to straightforwardly
calculate confidence regions for the acceleration observables by computing
$H(z)$, $D(z)$, $G(z)$ and the auxiliary observables 
$G_0(z)$, $f(z)G(z)$, and $\gamma(z)$ for each MCMC sample
using Eqs.~(\ref{eq:hz})$-$(\ref{eq:gamma_z}). The posterior distribution of
the model parameters $\bm{\theta}$ thus maps onto a distribution of each
acceleration observable at each redshift.  These redshift-dependent
distributions of the expansion and growth observables form the predictions
that we describe in the next section.

\section{Dark Energy Model Predictions}
\label{sec:predict}

In this section, we show the predictions for growth and expansion
observables from the combined current CMB, SN, BAO, and $H_0$
constraints. Since plotting full distributions for the six observables 
define in Sec.~\ref{sec:obs} at
several different redshifts is impractical, we instead plot only the regions
enclosing 68\% and 95\% of the models at each redshift, defined such that the number density of
models is equal at the upper and lower limit of each region.  
(When describing the predictions, we will typically quote the 68\% CL limits.)
To provide examples of
features of individual models that may not be apparent from the 68\% and 95\% CL limits,
we also plot the evolution of observables for the maximum
likelihood (ML) MCMC model within each model class. We caution, however, that the
MCMC algorithm is designed to approximate the overall shape of the likelihood
and is not optimized for precisely computing the ML
parameters, so the ``best fit'' models shown here may be slightly displaced
from the true ML points.

In most figures in this section, we compare the predictions for two model
classes, one of which is a subclass of the second, more general class (for
example, \lcdm\ and quintessence).  The potential to falsify the simpler class
in favor of the more complex one is greatest where the two sets of predictions
differ most, i.e.~where one class gives strong predictions and the other does
not.

\subsection{\lcdm}

\begin{figure}[tp]
\centerline{\psfig{file=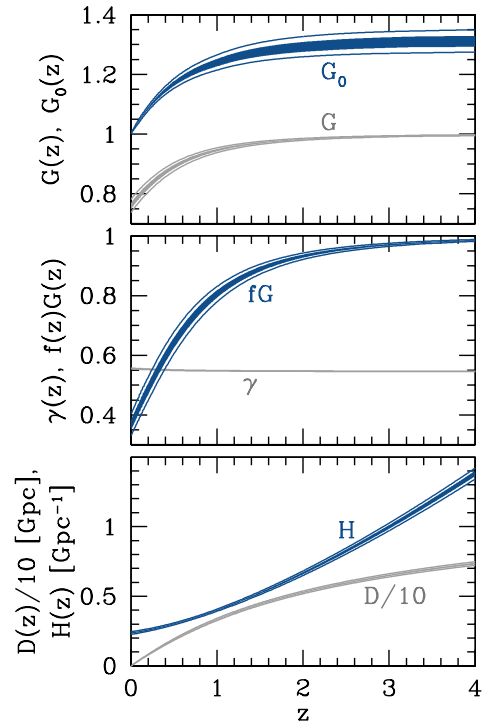, width=2.5in}}
\caption{Flat \lcdm\ predictions for growth and 
expansion observables, showing the 68\% CL (shading) and 95\% CL (curves) 
regions allowed by current CMB, SN, BAO, and $H_0$ data. 
Observables include the linear growth function normalized in two 
different ways, $G(z)$ equal to unity at high redshift and 
$G_0(z)=G(z)/G(0)$; the product of the differential growth rate 
and the growth function $f(z)G(z)$; 
the growth index $\gamma(z)$ which relates $f(z)$ and $\om(z)$; 
the expansion rate $H(z)$; and the comoving distance $D(z)$ 
(scaled by a factor of 1/10 in the lower panel).
Note that the separation between the 68\% and 95\% CL regions 
is not visible where the observables are extremely well predicted, 
e.g. in the $\gamma(z)$ predictions in the middle panel.
}
\label{fig:flcdm}
\end{figure}

We begin with the simplest and most predictive model class: flat $\Lambda$CDM.
Since $\ok=0$, this model has only two free dark energy parameters in
Eq.~(\ref{eq:parametersfull}), $\Omega_m$ and $\omhh$ (or $H_0$), 
providing very little freedom to alter the acceleration observables at {\it any} 
redshift as shown in Fig.~\ref{fig:flcdm}:
$H(z)$, $D(z)$, and $G(z)$ are currently predicted with a precision of $\sim 2\%$
(68\% CL) or better everywhere.   The velocity observable  $f(z)G(z)$ is predicted to better than 5\% and the growth index $\gamma$ to $0.1\%$.
These predictions are more precise than
current measurements of the acceleration observables at any redshift.

The strong predictions for flat \lcdm\ arise largely due to CMB constraints: the two parameters
$\om$ and $H_0$ are tied together by the measurement of $\omhh$,
and the remaining freedom in $H_0$ or the extragalactic distance scale is
fixed by the measurement of the distance to $z_*$.  
However, given the present uncertainties in $\omhh$ and $D(z_*)$, 
the addition of the other data (SN, BAO, and $H_0$) increases the 
precision of the predictions by almost a factor of 2 
relative to WMAP constraints alone.

The flat $\Lambda$CDM model is therefore highly falsifiable in that future measurements
may find that these quantities deviate substantially from the predictions. 
For example, an $H_0$ measurement with $\lesssim 2\%$ accuracy would match the precision of the
predictions and hence provide a sharp test of flat $\Lambda$CDM.   
These predictions are only a factor of $2-3$ weaker
than the Stage IV SN and CMB forecasts from MHH.  Since flat $\Lambda$CDM is the current
standard model of the cosmic expansion history and structure formation, falsifying it would represent
the most important observational breakthrough since the discovery of cosmic acceleration and would require revision of basic assumptions about the nature 
of dark energy, spatial curvature, or the theory of gravity.

\begin{figure}[tp]
\centerline{\psfig{file=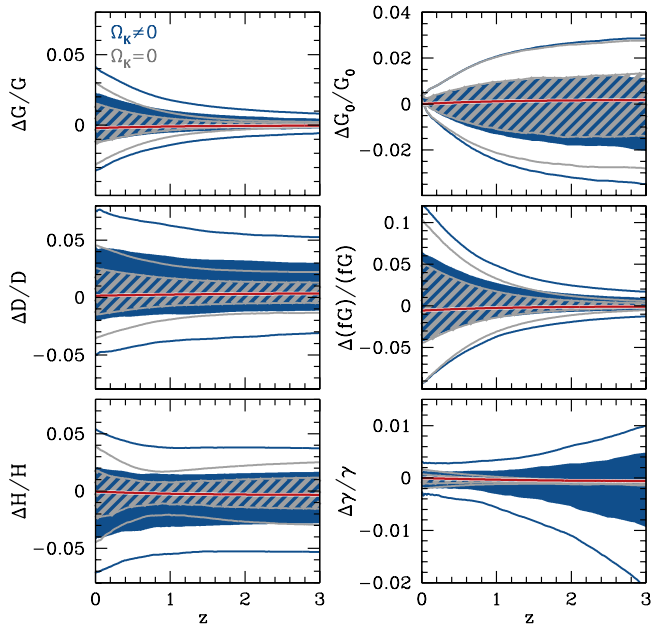, width=3.5in}}
\caption{Predicted growth and expansion observables for 
non-flat (dark blue) and flat (light gray) \lcdm, plotted 
relative to the reference cosmology (the best fit model for flat \lcdm).
Here and in subsequent figures, 
68\% CL regions are marked by shading, 95\% CL regions are bounded 
by solid curves, and red curves outlined in white show the 
best fit model of the more general (dark blue) model class 
(in this case, non-flat \lcdm). 
}
\label{fig:lcdm}
\end{figure}

Generalizing the model to $\Lambda$CDM with curvature increases the range of
predictions by less than a factor of~2.  In Fig.~\ref{fig:lcdm}, we plot the
predictions for flat and non-flat \lcdm\ relative to the ML flat $\Lambda$CDM
model with $\om=0.268$, $h=0.711$.  Curvature opens up the ability to free the
extragalactic distance scale from the constraints imposed by the CMB acoustic
peak measurements.  The tight constraints on SN, $H_0$, and BAO distances
limit this freedom.  Since the forecasts from MHH used only the current BAO
measurement and a weaker $H_0$ constraint as priors, the relative impact of
curvature here is substantially smaller.  In particular, predictions of the
growth function are nearly unchanged by curvature and still vary by less than
$2\%$.
Likewise, $fG$ is nearly unaffected by curvature.  Although
the growth index, $\gamma(z)$, is not as perfectly determined for non-flat
\lcdm, especially at high redshift, it is still predicted to better than 1\%
at $z \lesssim 3$, and both $D(z)$ and $H(z)$ are predicted to better than
$3\%$.  Any measurement that deviates by significantly more than these amounts
would prove that the dark energy is not a 
cosmological constant.\footnote{A substantial decrement in 
    growth from high redshifts,
    which in the context of our treatment would be interpreted as 
    evidence for early dark energy thus falsifying \lcdm,
    could alternately indicate neutrinos with more than the minimal 
    allowed masses.}

\subsection{Quintessence}

\begin{figure}[tp]
\centerline{\psfig{file=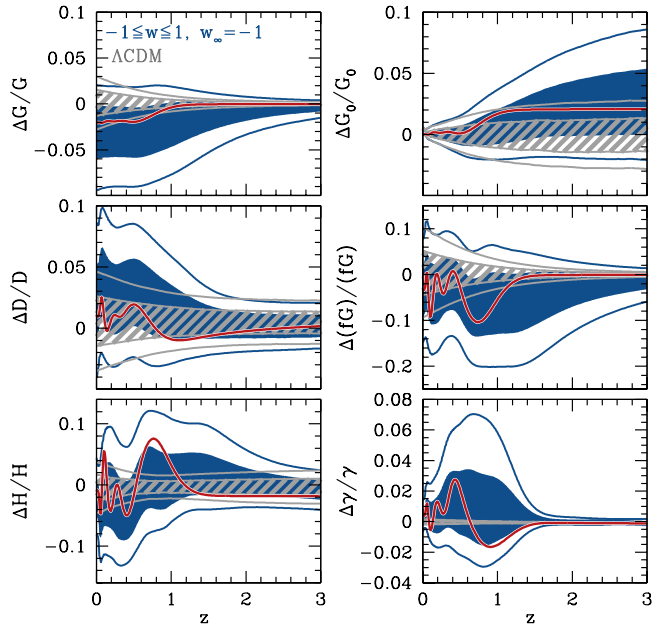, width=3.5in}}
\caption{Flat quintessence models without early dark energy (dark blue) 
vs.\ flat \lcdm\ (light gray). Other aspects here and in later figures follow
Fig.~\ref{fig:lcdm}.
}
\label{fig:quint}
\end{figure}

If $\Lambda$CDM is falsified, then in the context of dark energy 
we must consider models with $w(z) \ne -1$. Our
next class of models are therefore flat quintessence models 
with $w(z)$ parametrized by 10 principal components at $z<1.7$, assuming
no early dark energy (``$\winf=-1$''). The predictions for acceleration observables 
within this model class are compared with the flat \lcdm\ predictions
in Fig.~\ref{fig:quint}. 

Interestingly, the quintessence predictions are no longer centered on the
flat $\Lambda$CDM ML model.  From the $H(z)$ predictions which mainly reflect
variation in evolution of the dark energy density, we see that on average the
data favor a smaller low-redshift ($z\lesssim 0.5$) and larger
intermediate-redshift ($0.5\lesssim z\lesssim 2$) dark energy density.
Correspondingly, the best fit growth function $G(z)$ of $\Lambda$CDM is higher
than that of $\sim 85\%$ of the quintessence models in the chain.
Therefore a measurement of the growth relative to high redshift that is
smaller than the $\Lambda$CDM prediction by more than a few percent not only
rules out a cosmological constant but actually favors  these quintessence models.
The additional freedom in growth opens up predictions for $\gamma$
to include $2-3\%$ deviations at $z\lesssim 1$.

\begin{figure}[tp]
\centerline{\psfig{file=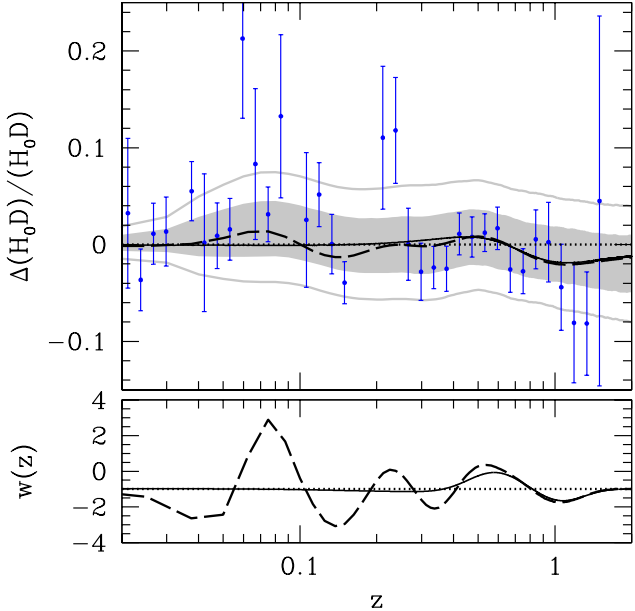, width=3.5in}}
\caption{Upper panel: Comparison of distance constraints from SN data and best
  fit models, plotted relative to the best fit $H_0D(z)$ for flat \lcdm\ (dotted
  line).  Blue points with error bars show the Union SN data in redshift
  bins of width $\Delta \log z = 0.05$. 
The best fit model for flat quintessence without
  early dark energy is plotted as a dashed curve, and the solid curve 
shows how the relative distances are affected by smoothing $w(z)$ 
for this model by a Gaussian of width $\sigma_z=0.1$. 
  The full distribution of relative distance predictions for this quintessence 
  model class is also
  shown with light gray shading (68\% CL) and curves (95\% CL).  
Lower panel: $w(z)$ for each of the models from the upper panel.  }
\label{fig:databf}
\end{figure}

Many of the shifts in the predictions relative to flat \lcdm\ are 
reflected in the evolution of $w(z)$ in the maximum likelihood model for 
flat quintessence without early dark energy. The ML model in this class marginally 
improves the fit to the current data sets relative to the 
\lcdm\ ML model, largely due to
variations in the SN data with redshift that are fit 
marginally better by dynamical dark energy than by a cosmological constant.
Figure~\ref{fig:databf} compares ML models, quintessence 
predictions, and relative distance constraints from the Union SN data sets 
at $z\lesssim 1$.
Freedom in $w(z)$ at these redshifts allows changes in the dark energy 
density to improve the fit to SN distances by $\dlnl\sim 4.5$.
However, some of this improvement is due to the large oscillations in 
the equation of state at $z\sim 0.1$, which are allowed to violate 
the $-1\leq w\leq 1$ bound due to the conservative implementation 
of the quintessence prior on PC amplitudes described in Sec.~\ref{sec:pcs}.
Smoothing the ML $w(z)$ by a Gaussian with width $\sigma_z\sim 0.1$ 
or requiring $w(z)$ to satisfy stricter quintessence bounds
reduces the improvement relative to \lcdm\ to $\dlnl\sim 2$, 
but has little effect on the overall distributions of the predicted 
observables.

Although differences in the ML models cause quintessence to not be centered
around \lcdm, the allowed {\it width} of quintessence predictions around
the maximum likelihood relative to \lcdm\ follows the expectations of the Stage IV predictions
from MHH except for being weaker by a factor of $2-3$.  The PCs allow for
oscillatory variations in $H(z)$, $f(z)G(z)$, and $\gamma(z)$ at $z<1$ that would
not be readily observable with expansion history or growth measures due to
limited resolution in redshift. On the other hand, $G(z)$, $G_0(z)$, and
$D(z)$ are still predicted with $\sim 2-3\%$ precision, so the class of flat
quintessence models without early dark energy remains highly falsifiable.

\begin{figure}[tp]
\centerline{\psfig{file=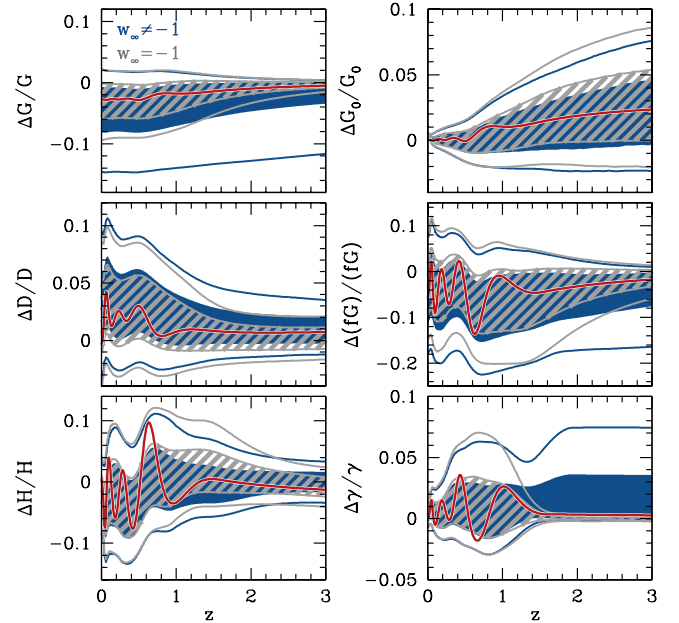, width=3.5in}}
\caption{Flat quintessence models with (dark blue) and without (light gray) 
early dark energy.
}
\label{fig:ede}
\end{figure}

Adding early dark energy to flat quintessence (Fig.~\ref{fig:ede}) 
has very little impact on the 68\% CL predictions of most observables due to the
restriction that $w\ge -1$ for a canonical scalar field.  To satisfy CMB
distance constraints, any increase in the expansion rate due to early dark
energy must be compensated by a lower expansion rate at intermediate redshift relative to
$z=0$, i.e. a dark energy density that decreases with increasing redshift
requiring $w<-1$.  While adding early dark energy does allow a larger
suppression of growth at high redshift (which is also a possible sign of 
massive neutrinos given current upper limits), a measurement of a $\gtrsim 10-15\%$
decrement or $\gtrsim 2\%$ increment in the growth relative to high redshift
would still suggest that a broader class of models is necessary.  This freedom in
growth leaves the amplitude relative to $z=0$ practically unchanged as the
$G_{0}(z)$ predictions show.  The only qualitative change with early dark energy
is to open up the allowed range in $\gamma(z)$ so that the high redshift end has as much
freedom as the low redshift end.  All of these trends for early dark energy
without curvature reflect those of the forecasts in MHH.

\begin{figure}[tp]
\centerline{\psfig{file=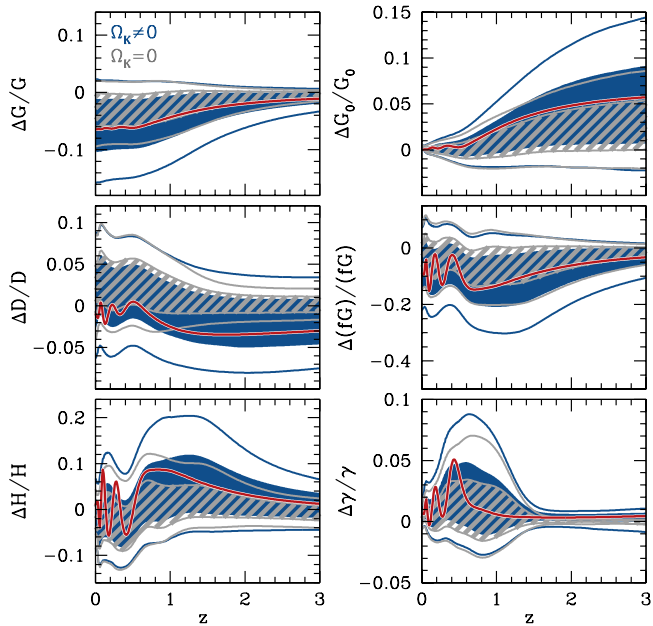, width=3.5in}}
\caption{Non-flat (dark blue) and flat (light gray) 
quintessence models without early dark energy.
}
\label{fig:curv}
\end{figure}

Including curvature in the quintessence class, but not early dark energy, opens up more
freedom as shown in Fig.~\ref{fig:curv}.  Now $z>2$ deviations in $D(z)$ are
allowed at the $\sim 5\%$ level relative to \lcdm.  
Thus a BAO distance measurement at $z>2$
 could falsify flat quintessence in favor of quintessence with curvature.
As discussed in MHH, because of the $w \ge -1$ quintessence bound, this additional freedom skews
 to smaller distances and lower growth relative to high redshift.

\begin{figure}[tp]
\centerline{\psfig{file=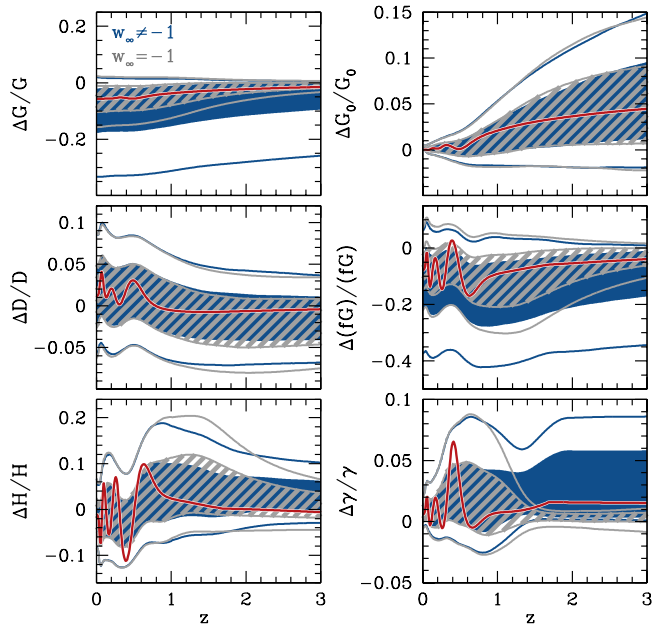, width=3.5in}}
\caption{Non-flat quintessence models with (dark blue)
and without (light gray) early dark energy.
}
\label{fig:curvede}
\end{figure}

Predictions from the most general quintessence class which includes 
both curvature and early dark energy, shown in Fig.~\ref{fig:curvede}, 
combine features of the previous quintessence classes in ways that 
are similar to the Stage IV predictions in MHH. 
The ML model in this class improves the fit to the combined data by 
$\dlnl\sim 4$, mostly due to changing the SN likelihood by $\dlnl\sim 5$;
however, removing the large low-$z$ oscillations by smoothing $w(z)$ reduces 
the improvement in the SN fit to $\dlnl\sim 2-3$.

The predictions for $G_0(z)$, $D(z)$, and $H(z)$, which 
were affected little by early dark energy alone, 
are nearly the same as those for non-flat 
quintessence without early dark energy.
The other observables show a mixture of the effects of
curvature at low $z$ and early dark energy at high $z$. Large suppression
($\gtrsim 20\%$) of $G(z)$ (and similarly $fG$) relative to \lcdm\ is allowed, but enhancement of
the growth function over the \lcdm\ best fit is still limited at the $\sim
2\%$ level.  Note that this upper limit on $G(z)$ is robust to neutrino mass 
uncertainties. Likewise, low-redshift distances (including $z_h
H_0^{-1}$) cannot be smaller than in $\Lambda$CDM by substantially more than
$\sim 2\%$.  As in Fig.~\ref{fig:ede}, 
the high redshift predictions for $\gamma(z)$ in Fig.~\ref{fig:curvede} weaken
substantially but only in the positive direction.  Indeed, all of the
observables display similar asymmetric weakening of the predictions with the
addition of curvature and early dark energy, which can be understood in terms
of the $w\geq -1$ quintessence bound.

The existence of an upper or lower bound on each observable that is robust to
freedom in curvature and early dark energy provides the possibility of
falsifying the entire quintessence model class.  In fact, in this most general
class, the statistical predictions from current SN and CMB bounds are already
comparable to those that can be achieved by a Stage IV version of these
probes, which can be understood from the fact
 that the forecasts from MHH used current BAO and
$H_0$ measurements.

The comparable predictions in large part reflect the fact that curvature is
already well constrained through the BAO and $H_0$ measurements.  
The constraint in this most
general class of quintessence models is $-0.006<\Omega_K<0.033$ (95\%~CL), 
a factor of $\sim 2$ weaker than for non-flat \lcdm\ and skewed
toward open models due to the quintessence prior on $w(z)$.

\begin{figure}[tp]
\centerline{\psfig{file=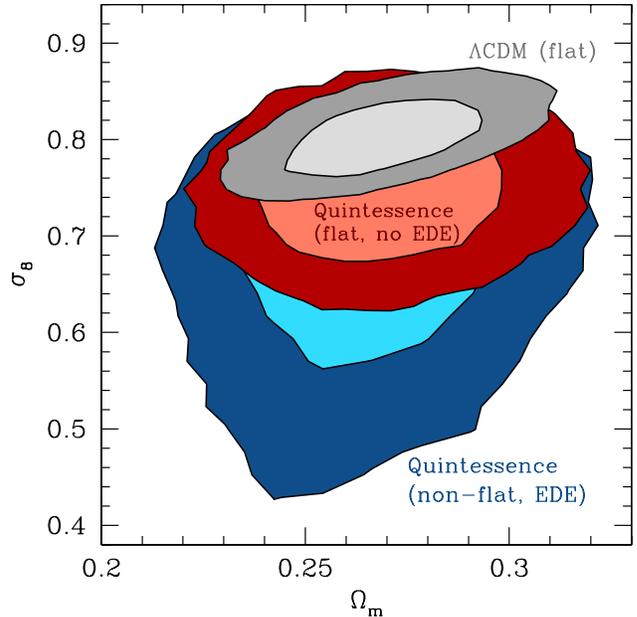, width=3.5in}}
\caption{Predictions for $\sigma_8$ and $\om$ for flat \lcdm\ 
(gray contours, top), flat quintessence without early dark energy
(red contours, middle), 
 and non-flat quintessence with early dark energy 
(blue contours, bottom), showing 68\% CL (light) and 95\% CL (dark) regions.
}
\label{fig:sigma8}
\end{figure}

Finally, as an example of the use of the asymmetric quintessence predictions,
we consider the application of these results 
to observables which measure some combination of $\sigma_8$ and $\om$.  
To compute predictions for $\sigma_8$ given our predictions for the raw acceleration observables, 
we use the fitting formula \cite{Hu_Jain}
\begin{eqnarray}
\sigma_8 &=& \frac{G(z=0)}{0.76}\left[\frac{A_s(k=0.05~{\rm Mpc}^{-1})}{3.12\times 10^{-9}}\right]^{1/2} \left(\frac{\obhh}{0.024}\right)^{-1/3} \nonumber \\
&& \times \left(\frac{\omhh}{0.14}\right)^{0.563}
\left(\frac{h}{0.72}\right)^{0.693}(3.123h)^{(n_s-1)/2}
\label{eq:sigma8}
\end{eqnarray}
for each model sampled in the MCMC likelihood analysis. 
Note that on top of allowed
variations in $G(z=0)$, $\sigma_{8}$ predictions include uncertainties in the
reionization optical depth $\tau$ through its covariance with $A_{s}$. While
this analysis assumes instantaneous reionization, the uncertainty introduced
by more general ionization histories is small \cite{MorHu08}.
We have checked that the $\sigma_8$ distributions obtained using 
Eq.~(\ref{eq:sigma8}) closely match those 
from the more accurate computation of $\sigma_8$ using CAMB.
The joint predictions for $\sigma_8$ and $\om$ from the current SN, CMB, 
BAO, and $H_0$ constraints are shown in Fig.~\ref{fig:sigma8} for 
flat \lcdm\ and two quintessence model classes.

In particular, in the context of flat \lcdm\ the current SN, CMB, BAO, and
$H_0$ data predict the combination best measured by the local abundance of
massive galaxy clusters to be $0.394 < \sigma_8 \om^{0.5} < 0.441$ (68\% CL).
Flat quintessence without early dark energy weakens the lower end somewhat but
leaves the upper limit nearly unchanged: $0.358 < \sigma_8 \om^{0.5} < 0.419$.
Quintessence with both early dark energy and curvature yields $0.306 < \sigma_8
\om^{0.5} < 0.396$.  Therefore a measurement of a local cluster abundance in
significant excess of the flat \lcdm\ predictions rules out the whole quintessence class,
whereas a measurement that is substantially lower would remain consistent with
quintessence but would rule out a cosmological constant (see also \cite{Kunz_sigma8}).
A measurement below the flat \lcdm\ prediction by $\lesssim 10\%$ could 
also indicate large neutrino masses, but an excess cluster abundance could 
not be alternately explained by massive neutrinos.
Current cluster surveys, with $\sim 5\%$ measurements of similar 
combinations of $\sigma_8$ and $\om$ \cite{Vikhlinin,Rozo,Mantz},
are beginning to reach the precision necessary to test these predictions.  
In fact, the
lack of an observed excess already places strong constraints on modified
gravity explanations of cosmic acceleration \cite{Schmidt:2009am}.

\subsection{Smooth $w_0-w_a$ Dark Energy}

As a final case we consider the class of models defined by an equation of
state $w(z)= w_{0} +(1-a)w_{a}$ \cite{Chevallier_Polarski,Linder_wa} under the
assumption that dark energy is smooth relative to matter.
Unlike our previous cases, this class does not define a physical candidate for
dark energy such as the cosmological constant or a scalar field but rather
represents a simple but illustrative phenomenological parametrization.  Note
that early dark energy is included in this parametrization since 
$\lim_{z\to\infty} w(z) = w_0 + w_a$.
 
The predictions for the $w_0-w_a$ model class serve two purposes.  First, the
comparison of predictions for smooth, monotonic $w_0-w_a$ models with 
those for PC quintessence models test the dependence of the
predictions on rapid transitions and non-monotonic evolution of the
equation of state.  The second use of the $w_0-w_a$ predictions is to
illustrate how predictions are affected by the $-1\leq w(z) \leq 1$
quintessence bound.  Unlike the model classes where $w(z)$ is parametrized by
principal components, it is simple to impose a strict 
quintessence prior on $w_0-w_a$ models
by requiring $-1\leq w_0 \leq 1$ and $-1\leq w_0+w_a\leq 1$.  We compare
predictions using this prior with the more general case, where the priors are
weak enough that constraints on $w_0$ and $w_a$ are determined solely 
by the data (``no $w$ prior'').

\begin{figure}[tp]
\centerline{\psfig{file=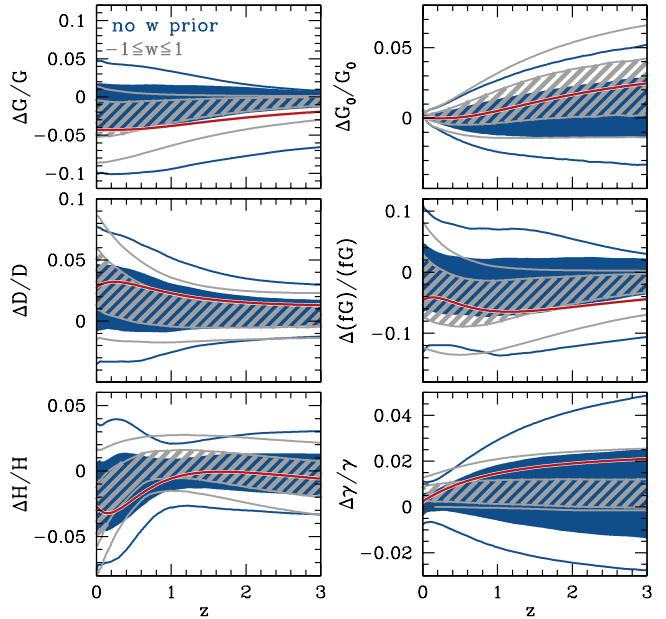, width=3.5in}}
\caption{Flat $w_0-w_a$ without priors on $w(z)$ (dark blue) and with
  quintessence priors ($-1\leq w_0\leq 1$, $-1\leq w_0+w_a\leq 1$; light
  gray).  
}
\vskip 0.25cm
\label{fig:w0waf}
\end{figure}

\begin{figure}[tp]
\centerline{\psfig{file=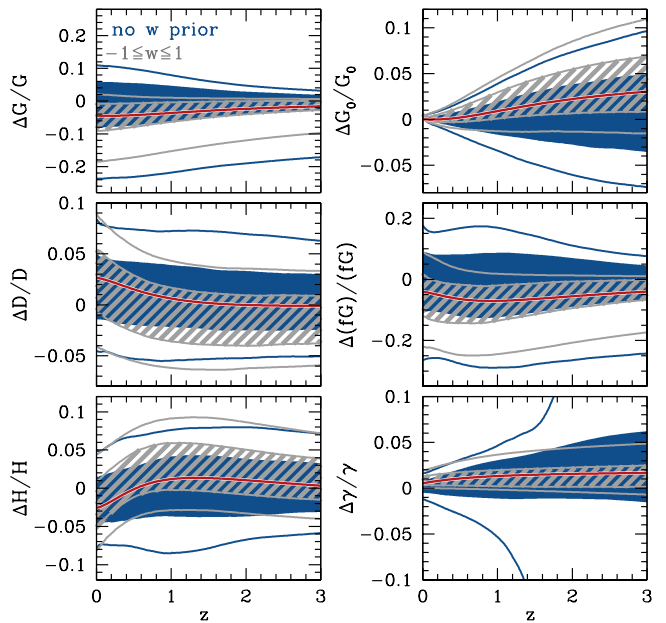, width=3.5in}}
\caption{Non-flat $w_0-w_a$ without priors on $w(z)$ (dark blue) and 
with quintessence priors ($-1\leq w_0\leq 1$, $-1\leq w_0+w_a\leq 1$; 
light gray).
}
\vskip 0.25cm
\label{fig:w0wac}
\end{figure}

A fair comparison can be made between the predictions for flat and non-flat
$w_{0}-w_{a}$ models with the $-1\leq w\leq 1$ prior (light gray contours in
Figs.~\ref{fig:w0waf} and \ref{fig:w0wac}) and PC quintessence models with
early dark energy (dark blue contours in Figs.~\ref{fig:ede} and
\ref{fig:curvede}).  In particular, observables relatively insensitive to both
the amount of early dark energy and large changes in the PC equation of
state at low redshift, such as $G_0(z)$ and $D(z)$, 
are generally in good agreement.
The expansion rate and growth rate are more sensitive to sudden changes in
$w(z)$ than the distances and the integrated growth function. Therefore, the
impact of large, low-$z$ oscillations in the PCs is greatest for $H(z)$, 
$f(z)G(z)$, and $\gamma(z)$ at $z\lesssim 1$, increasing the width of those predictions
relative to the corresponding predictions for the smooth $w_0-w_a$ models. The
PC quintessence models also have more freedom in early dark energy than
$w_0-w_a$ models since $\winf$, unlike $w_0+w_a$, is completely free from the
low-redshift SN, BAO, and $H_0$ constraints. As a result, $w_0-w_a$
predictions for $G(z)$ and the high-redshift values of $\gamma(z)$ 
and $f(z)G(z)$ are stronger than, but still qualitatively similar to, 
those for PC quintessence with early dark energy.

Like the PC quintessence predictions, the predictions for $w_0-w_a$ models 
bounded by $-1\leq w\leq 1$ are shifted relative to flat \lcdm\ 
due to marginal improvements in the fit to SN data ($\dlnl\sim 0.5$ for 
the ML model) enabled by an evolving equation of state.
This is a somewhat smaller change in the likelihood than 
for PC quintessence models, but the  
magnitude of the ML model shift in the observables is similar for 
$w_0-w_a$ and PC quintessence, at least for those observables that 
depend little on early dark energy.

Comparing the two sets of predictions in Figs.~\ref{fig:w0waf}
and~\ref{fig:w0wac} (no $w$ prior vs.\ the $-1\leq w\leq 1$ prior) shows the
effect on the $w_0-w_a$ predictions of allowing freedom in $w(z)$ beyond that
allowed by the quintessence bounds.  As discussed in MHH, eliminating these
bounds makes the range in predictions for observables such as growth more
symmetric around the best fit for flat \lcdm\ since $w(z)$ is allowed to cross
below $w=-1$.  In particular, growth in excess of flat
  $\Lambda$CDM is now allowed.  Based on the analysis of MHH, we expect the
  amount of the remaining skewness in the predictions around flat $\Lambda$CDM
  to be affected by the available volume of parameter space as determined by
  how priors on dark energy parameters weight models with $w<-1$ relative to
  those with $w>-1$.

Removing the quintessence bounds also allows models with greater amounts of
early dark energy, and (for non-flat $w_0-w_a$) more closed models, to fit the
data.  A notable consequence for models with nonzero curvature is that the
predictions for $\gamma(z)$ at 95\% CL diverge at $z>1$.  
This is the same effect noted in MHH for $\gamma(z)$ forecasts
in the non-flat smooth dark energy model class.
The divergence in the tails of the high-redshift $\gamma(z)$ distribution 
is caused by the appearance of
a singularity in $\gamma(z)$ for closed models where $\ok$ is sufficiently
negative so that $\om(z)$ crosses unity at some redshift; when $\om(z)=1$,
$\gamma(z)$ is no longer well defined by Eq.~(\ref{eq:gamma_z}).  Such caveats
must be kept in mind when using $\gamma$ as a test of not only quintessence
but of all smooth dark energy models.

\section{Discussion}
\label{sec:discussion}

Any given class of dark energy models makes concrete predictions for the
relationship between the expansion history, geometry, and growth 
of structure as a function
of redshift. Therefore, current distance-based measurements, though limited in
redshift, make predictions for other dark energy observables that can be used
to test and potentially rule out whole classes of dark energy models.

In this paper we present the allowed ranges for the expansion rate $H(z)$,
distances $D(z)$, the linear growth rate $G(z)$, 
and several auxiliary growth observables from the current
combination of cosmological measurements of supernovae, the cosmic microwave
background, baryon acoustic oscillations, and the Hubble constant.  In
particular, growth at any redshift or a Hubble constant in significant excess
of $2\%$ ($68\%$ CL range) of the current best fit $\Lambda$CDM model would
falsify both a cosmological constant and more general quintessence 
models with or without curvature and early dark energy.  On the
other hand, comparable measurements of a decrement in these quantities would
rule out a cosmological constant but would be fully consistent
with quintessence.  Alternately, a substantial reduction in growth 
relative to the expectation for \lcdm\ could indicate neutrinos with 
large masses ($\sum m_{\nu}>0.05$~eV).

Remarkably, predictions for the main acceleration observables, $H(z)$, $D(z)$,
and $G(z)$, are only weaker than Stage IV SN and CMB predictions (MHH) by a
factor of $\sim 2-3$.  However, this improvement applies across a wide range
of redshifts, indicating that multiple phenomenological parameters may each be
improved by this factor.  For example, parameter-based figures of merit
effectively involve products of individual parameters (e.g.\ area in the
$w_0-w_a$ plane \cite{Huterer_Turner,DETF} or volume of the principal
component parameter error ellipsoid \cite{Albrecht_Bernstein,FoMSWG}), and in
such figures of merit the total improvement with future data can be
significant.  
If novel dark energy physics affects small pockets of these high-dimensional
parameter spaces --- that is, if only specific dark energy parameter combinations
are sensitive to new physics --- then these multiparameter figures of merit
will justly indicate a much more significant improvement with future cosmological
data.

In this work we have considered only known and quantifiable sources of error
in the current data.  Recent analyses of supernova data (e.g.~\cite{Constitution,SDSS_SN,Kelly09}) indicate that unknown systematic
errors remain and can significantly affect cosmological constraints.
Furthermore, the systematic error estimates used here for the SN data were
optimized for models with a cosmological constant and therefore may be
underestimated for dynamical dark energy \cite{SCP_Union}.  We intend to
explore the implications of SN systematics for dark energy predictions in
future work.  Our predictive methodology can alternately be viewed as a means
of ferreting out unknown systematics by looking for inconsistencies between
the predictions from one set of observations and data from another.

Over the course of this study, new data have become available that could
improve the predictions for acceleration observables or begin to test
predictions within the various classes. In particular, BAO measurements from
SDSS DR7 and 2dFGRS provide a 2.7\% constraint on $D_V(z=0.275)$ and a 3.7\%
constraint on $D_V(z=0.35)/D_V(z=0.2)$ \cite{Percival09}.  We have estimated the
impact of these new measurements on our predictions by using the updated BAO
likelihood to modify the weighting of MCMC samples for each model class.  For
all quintessence model classes, the effect of updating the BAO data is
negligible for most observables except for $D(z\lesssim 1)$ and (to a lesser
extent) $H(z\lesssim 0.5)$, reflecting the improved BAO constraint on
low-redshift $D$ and $H$.

The impact of the newer BAO measurements on \lcdm\ models is greater than
for quintessence since the reduced freedom in dark energy evolution ties
low-redshift measurements to high-redshift predictions.  The updated BAO
constraints exclude models on one side of the predicted observable
distributions in Fig.~\ref{fig:lcdm}, reducing their width by $10-30\%$ and
shifting the distributions by an equal amount.  However, these changes appear
to be mainly due to a slight tension between the new BAO constraints and the
other data sets used for \lcdm\ predictions. Note that the BAO constraints of
Ref.~\cite{Percival09} are still less precise than the flat \lcdm\ predictions
in Fig.~\ref{fig:lcdm} and comparable to the non-flat \lcdm\ predictions, so
they do not yet represent a significant additional test of the cosmological
constant.

Falsifiable predictions from current data reveal many opportunities for sharp
observational tests of paradigms for cosmic acceleration by requiring 
consistency within a given theoretical framework between observables 
that depend on the expansion history, geometry, and growth of structure 
in the universe.  These predictions can be used to 
inform future surveys as to the optimal choice of observables, redshifts, and
required measurement accuracies for testing whole
classes of dark energy models.
Falsification of even the simplest model, flat \lcdm, would have 
revolutionary consequences for cosmology and fundamental physics.

\vspace{1cm}
{\it Acknowledgments:} We thank David Weinberg for useful conversations about this
work. MM and WH were supported by the KICP under NSF contract PHY-0114422.  MM
was additionally supported by the NSF GRFP and CCAPP at Ohio State; WH by DOE
contract DE-FG02-90ER-40560 and the Packard Foundation; DH by the DOE OJI
grant under contract DE-FG02-95ER40899, NSF under contract AST-0807564, and
NASA under contract NNX09AC89G.

\vfill
\bibliographystyle{arxiv_physrev}
\bibliography{pccurrent}

\def\eprinttmppp@#1arXiv:@{#1}
\providecommand{\arxivlink[1]}{\href{http://arxiv.org/abs/#1}{arXiv:#1}}
\providecommand{\arxivlinknopre[1]}{\href{http://arxiv.org/abs/#1}{#1}}
\providecommand{\eprintmod}[1][XXXX.XXXX]{\IfSubStr{#1}{arXiv}{\arxivlinknopre%
{#1}}{\arxivlink{#1}}}
\providecommand{\adsurl}[1]{\href{#1}{ADS}}
\begin{thebibliography}{62}
\expandafter\ifx\csname natexlab\endcsname\relax\def\natexlab#1{#1}\fi
\expandafter\ifx\csname bibnamefont\endcsname\relax
  \def\bibnamefont#1{#1}\fi
\expandafter\ifx\csname bibfnamefont\endcsname\relax
  \def\bibfnamefont#1{#1}\fi
\expandafter\ifx\csname citenamefont\endcsname\relax
  \def\citenamefont#1{#1}\fi
\expandafter\ifx\csname url\endcsname\relax
  \def\url#1{\texttt{#1}}\fi
\expandafter\ifx\csname urlprefix\endcsname\relax\def\urlprefix{URL }\fi

\bibitem{Spergel_2003}
D.~N. Spergel {\em et~al.},
\newblock Astrophys. J. Suppl. {\bf 148}, 175 (2003),
  [\eprintmod[astro-ph/0302209]].

\bibitem{Hu_standards}
W.~Hu,
\newblock ASP Conf. Ser. {\bf 339}, 215 (2005), [\eprintmod[astro-ph/0407158]].

\bibitem{PaperI}
M.~J. {Mortonson}, W.~{Hu} and D.~{Huterer},
\newblock Phys. Rev. {\bf D79}, 023004 (2009), [\eprintmod[0810.1744]],
\newblock (MHH).

\bibitem{DETF}
A.~Albrecht {\em et~al.},
\newblock \eprintmod[astro-ph/0609591].

\bibitem{Huterer_Cooray}
D.~Huterer and A.~Cooray,
\newblock Phys. Rev. {\bf D71}, 023506 (2005), [\eprintmod[astro-ph/0404062]].

\bibitem{Wang_Tegmark_2005}
Y.~Wang and M.~Tegmark,
\newblock Phys. Rev. {\bf D71}, 103513 (2005), [\eprintmod[astro-ph/0501351]].

\bibitem{Riess_2006}
A.~G. Riess {\em et~al.},
\newblock Astrophys. J. {\bf 659}, 98 (2007), [\eprintmod[astro-ph/0611572]].

\bibitem{Zunckel_Trotta}
C.~Zunckel and R.~Trotta,
\newblock Mon. Not. Roy. Astron. Soc. {\bf 380}, 865 (2007),
  [\eprintmod[astro-ph/0702695]].

\bibitem{Sullivan_Cooray_Holz}
S.~Sullivan, A.~Cooray and D.~E. Holz,
\newblock JCAP {\bf 0709}, 004 (2007), [\eprintmod[0706.3730]].

\bibitem{Zhao_Huterer_Zhang}
G.-B. Zhao, D.~Huterer and X.~Zhang,
\newblock Phys. Rev. {\bf D77}, 121302 (2008), [\eprintmod[0712.2277]].

\bibitem{Zhao_Zhang:2009}
G.-B. Zhao and X.~Zhang,
\newblock \eprintmod[0908.1568].

\bibitem{Serra:2009}
P.~Serra {\em et~al.},
\newblock \eprintmod[0908.3186].

\bibitem{Kujat}
J.~Kujat, A.~M. Linn, R.~J. Scherrer and D.~H. Weinberg,
\newblock Astrophys. J. {\bf 572}, 1 (2002), [\eprintmod[astro-ph/0112221]].

\bibitem{Chongchitnan_Efstathiou}
S.~Chongchitnan and G.~Efstathiou,
\newblock Phys. Rev. {\bf D76}, 043508 (2007), [\eprintmod[0705.1955]].

\bibitem{Sahlen05}
M.~Sahlen, A.~R. Liddle and D.~Parkinson,
\newblock Phys. Rev. {\bf D72}, 083511 (2005), [\eprintmod[astro-ph/0506696]].

\bibitem{Huterer_Peiris}
D.~Huterer and H.~V. Peiris,
\newblock Phys. Rev. {\bf D75}, 083503 (2007), [\eprintmod[astro-ph/0610427]].

\bibitem{ZhaKnoTys08}
H.~Zhan, L.~Knox and J.~A. Tyson,
\newblock Astrophys. J. {\bf 690}, 923 (2009), [\eprintmod[0806.0937]].

\bibitem{nu_oscill}
U.~{Dore} and D.~{Orestano},
\newblock Reports on Progress in Physics {\bf 71}, 106201 (2008),
  [\eprintmod[0811.1194]].

\bibitem{Reid_SDSSDR7}
B.~A. Reid {\em et~al.},
\newblock \eprintmod[0907.1659].

\bibitem{Percival_White}
W.~J. {Percival} and M.~{White},
\newblock \mnras {\bf 393}, 297 (2009), [\eprintmod[0808.0003]].

\bibitem{WangSteinhardt}
L.~{Wang} and P.~J. {Steinhardt},
\newblock \apj {\bf 508}, 483 (1998), [\eprintmod[astro-ph/9804015]].

\bibitem{Linder_gamma}
E.~V. {Linder},
\newblock Phys. Rev. {\bf D72}, 043529 (2005), [\eprintmod[astro-ph/0507263]].

\bibitem{SCP_Union}
M.~Kowalski {\em et~al.},
\newblock Astrophys. J. {\bf 686}, 749 (2008), [\eprintmod[0804.4142]].

\bibitem{Union_like}
\url{http://supernova.lbl.gov/Union/}.

\bibitem{Komatsu_2008}
E.~Komatsu {\em et~al.},
\newblock Astrophys. J. Suppl. {\bf 180}, 330 (2009), [\eprintmod[0803.0547]].

\bibitem{Nolta_2008}
M.~R. Nolta {\em et~al.},
\newblock Astrophys. J. Suppl. {\bf 180}, 296 (2009), [\eprintmod[0803.0593]].

\bibitem{Dunkley_2008}
J.~Dunkley {\em et~al.},
\newblock Astrophys. J. Suppl. {\bf 180}, 306 (2009), [\eprintmod[0803.0586]].

\bibitem{WMAP_like}
\url{http://lambda.gsfc.nasa.gov/}.

\bibitem{Lewis:1999bs}
A.~Lewis, A.~Challinor and A.~Lasenby,
\newblock Astrophys. J. {\bf 538}, 473 (2000), [\eprintmod[astro-ph/9911177]].

\bibitem{camb_url}
\url{http://camb.info/}.

\bibitem{PPF}
W.~{Fang}, W.~{Hu} and A.~{Lewis},
\newblock Phys. Rev. {\bf D78}, 087303 (2008), [\eprintmod[0808.3125]].

\bibitem{ppf_url}
\url{http://camb.info/ppf/}.

\bibitem{Eisenstein}
D.~J. Eisenstein {\em et~al.},
\newblock Astrophys. J. {\bf 633}, 560 (2005), [\eprintmod[astro-ph/0501171]].

\bibitem{Percival09}
W.~J. {Percival} {\em et~al.},
\newblock \mnras  (2009), [\eprintmod[0907.1660]].

\bibitem{SHOES}
A.~G. Riess {\em et~al.},
\newblock Astrophys. J. {\bf 699}, 539 (2009), [\eprintmod[0905.0695]].

\bibitem{HubTrans}
M.~Mortonson, W.~Hu and D.~Huterer,
\newblock Phys. Rev. {\bf D80}, 067301 (2009), [\eprintmod[0908.1408]].

\bibitem{Huterer_Starkman}
D.~Huterer and G.~Starkman,
\newblock Phys. Rev. Lett. {\bf 90}, 031301 (2003),
  [\eprintmod[astro-ph/0207517]].

\bibitem{Hu_PC}
W.~Hu,
\newblock Phys. Rev. {\bf D66}, 083515 (2002), [\eprintmod[astro-ph/0208093]].

\bibitem{SNAP}
G.~Aldering {\em et~al.},
\newblock \eprintmod[astro-ph/0405232].

\bibitem{Samsing_Linder}
J.~Samsing and E.~V. Linder,
\newblock \eprintmod[0908.2637].

\bibitem{Christensen:2001gj}
N.~Christensen, R.~Meyer, L.~Knox and B.~Luey,
\newblock Class. Quant. Grav. {\bf 18}, 2677 (2001),
  [\eprintmod[astro-ph/0103134]].

\bibitem{Kosowsky:2002zt}
A.~Kosowsky, M.~Milosavljevic and R.~Jimenez,
\newblock Phys. Rev. {\bf D66}, 063007 (2002), [\eprintmod[astro-ph/0206014]].

\bibitem{Dunetal05}
J.~{Dunkley}, M.~{Bucher}, P.~G. {Ferreira}, K.~{Moodley} and C.~{Skordis},
\newblock \mnras {\bf 356}, 925 (2005), [\eprintmod[astro-ph/0405462]].

\bibitem{Lewis:2002ah}
A.~Lewis and S.~Bridle,
\newblock Phys. Rev. {\bf D66}, 103511 (2002), [\eprintmod[astro-ph/0205436]].

\bibitem{cosmomc_url}
\url{http://cosmologist.info/cosmomc/}.

\bibitem{gelman/rubin}
A.~Gelman and D.~Rubin,
\newblock Statistical Science {\bf 7}, 452 (1992).

\bibitem{Hu_Jain}
W.~Hu and B.~Jain,
\newblock Phys. Rev. {\bf D70}, 043009 (2004), [\eprintmod[astro-ph/0312395]].

\bibitem{MorHu08}
M.~J. {Mortonson} and W.~{Hu},
\newblock \apj {\bf 672}, 737 (2008), [\eprintmod[0705.1132]].

\bibitem{Kunz_sigma8}
M.~Kunz, P.~S. Corasaniti, D.~Parkinson and E.~J. Copeland,
\newblock Phys. Rev. {\bf D70}, 041301 (2004), [\eprintmod[astro-ph/0307346]].

\bibitem{Vikhlinin}
A.~Vikhlinin {\em et~al.},
\newblock Astrophys. J. {\bf 692}, 1060 (2009), [\eprintmod[0812.2720]].

\bibitem{Rozo}
E.~Rozo {\em et~al.},
\newblock \eprintmod[0902.3702].

\bibitem{Mantz}
A.~Mantz, S.~W. Allen, D.~Rapetti and H.~Ebeling,
\newblock \eprintmod[0909.3098].

\bibitem{Schmidt:2009am}
F.~Schmidt, A.~Vikhlinin and W.~Hu,
\newblock Phys. Rev. {\bf D80}, 083505 (2009), [\eprintmod[0908.2457]].

\bibitem{Chevallier_Polarski}
M.~Chevallier and D.~Polarski,
\newblock Int. J. Mod. Phys. {\bf D10}, 213 (2001),
  [\eprintmod[gr-qc/0009008]].

\bibitem{Linder_wa}
E.~V. Linder,
\newblock Phys. Rev. Lett. {\bf 90}, 091301 (2003),
  [\eprintmod[astro-ph/0208512]].

\bibitem{Huterer_Turner}
D.~Huterer and M.~S. Turner,
\newblock Phys. Rev. {\bf D64}, 123527 (2001), [\eprintmod[astro-ph/0012510]].

\bibitem{Albrecht_Bernstein}
A.~J. Albrecht and G.~Bernstein,
\newblock Phys. Rev. {\bf D75}, 103003 (2007), [\eprintmod[astro-ph/0608269]].

\bibitem{FoMSWG}
A.~J. Albrecht {\em et~al.},
\newblock \eprintmod[0901.0721].

\bibitem{SDSS_SN}
R.~Kessler {\em et~al.},
\newblock Astrophys. J. Suppl. {\bf 185}, 32 (2009), [\eprintmod[0908.4274]].

\bibitem{Kelly09}
P.~L. Kelly, M.~Hicken, D.~L. Burke, K.~S. Mandel and R.~P. Kirshner,
\newblock \eprintmod[0912.0929].

\bibitem{Constitution}
M.~Hicken {\em et~al.},
\newblock Astrophys. J. {\bf 700}, 1097 (2009), [\eprintmod[0901.4804]].

\end{thebibliography}

\end{document}